\documentclass[aps,prd,twocolumn,superscriptaddress]{revtex4}
\usepackage{graphicx}
\usepackage[hypertex]{hyperref}

\def\beq{\begin{equation}}
\def\eeq{\end{equation}}
\def\bea{\begin{eqnarray}}
\def\eea{\end{eqnarray}}

\begin{document}

\title{Cuscuton and low energy limit of Ho${\rm{\bf \check{r}}}$ava-Lifshitz gravity}

\author{Niayesh Afshordi}\affiliation{Perimeter Institute
for Theoretical Physics, 31 Caroline St. N., Waterloo, ON, N2L 2Y5, Canada}

\date{\today}
\preprint{astro-ph/yymmnnn}

\begin{abstract}
A proposal for a power-counting renormalizable theory of quantum gravity at a Lifshitz point was recently put forth by Ho${\rm \check{r}}$ava \cite{Horava:2009uw}, and has been since dubbed as Ho${\rm \check{r}}$ava-Lifshitz gravity. The theory explicitly breaks Lorentz invariance, which introduces an apparent extra scalar degree of freedom. In this note, we show that the low energy limit of (non-projectible) Ho${\rm \check{r}}$ava-Lifshitz gravity is uniquely given by the quadratic cuscuton model: a covariant scalar field theory with an infinite speed of sound and a quadratic potential, which is minimally coupled to Einstein gravity. This implies that the extra scalar is non-dynamical to all orders in perturbation theory. Using current cosmological constraints on the quadratic cuscuton model, we can constrain the low energy Lorentz breaking parameter of Ho${\rm \check{r}}$ava-Lifshitz theory (which leads to a running of Planck mass on the Hubble scale) to $|\lambda-1| < 0.014$ (at 95\% confidence level). We also point out that, with reasonable boundary conditions, the spatial hypersurfaces in this theory are constant mean curvature (CMC) or uniform expansion surfaces at low energies, and introduce geometrical symmetries that can protect the non-dynamical nature of these theories from quantum corrections. We also notice that the theory with $\lambda < 1$ might suffer from a non-perturbative UV instability.

\end{abstract}

\maketitle

\section{Introduction}

A consistent quantization of the theory of gravity has been among the biggest challenges in theoretical physics over the past century.
An intriguing possibility that might explain why a satisfactory solution to this problem has evaded physicists for so long is that Lorentz symmetry, which is the benchmark of Einstein's theory of relativity, is only a low-energy emergent phenomenon, and the fundamental theory has a preferred reference frame. An explicit example of this possibility is a power-counting  renormalizable theory of gravity which includes higher spatial (but not time) derivatives of the 3-curvature tensor \cite{Horava:2009uw} (also see \cite{Moffat:1992bf,Moffat:1992ud}). The theory does not suffer from any ghosts (at least in its tensor degrees of freedom), but at the expense of maximal breaking of Lorentz symmetry (or the so-called $z=3$ Lifshitz point) at high energies. The remaining symmetries of the theory are only arbitrary {\it spatial} diffeomorphisms and (space-independent) time reparametrizations.

In this note, we show that the low-energy limit of the Ho${\rm \check{r}}$ava-Lifshitz theory reduces to Einstein's gravity, minimally coupled to cuscuton field theory: a scalar field theory with an infinite speed of sound and a quadratic potential \cite{Afshordi:2006ad,Afshordi:2007yx}. The scalar field is non-dynamical and the theory has a large class of geometric symmetries that can protect this behavior from quantum corrections. This sheds new light onto Lorentz-violating UV-completions of the theory of gravity, that do not have new dynamical degrees of freedom or ghost-like catastrophes.

\section{Equivalence of cuscuton and Ho${\bf \check{R}}$ava-Lifshitz gravities}
Within the reduced symmetry class of the theory, the most general  low-energy limit of (non-projectible) Ho${\rm \check{r}}$ava-Lifshitz gravity action is particularly interesting:
\beq
S_{HL} = S_{EH}+ \frac{1-\lambda}{16\pi G_N}\int d^4x \sqrt{-g} K^2,
\eeq
where $S_{EH}$ is the Einstein-Hilbert action of General Relativity, while $K$ is the mean extrinsic curvature of constant time hypersurfaces in the Ho${\rm \check{r}}$ava-Lifshitz theory, i.e. in a general coordinate system we have:
\beq
K \equiv \nabla_{\mu}u^{\mu},  u^\mu = \frac{\partial^{\mu}\psi}{\sqrt{\partial^{\nu}\psi\partial_{\nu}\psi}},
\eeq
where $\psi$ is the coordinate time in the preferred frame of the Ho${\rm \check{r}}$ava-Lifshitz theory. Therefore, the correction to Einstein-Hilbert action at low energies is given by:
\bea
S_{\lambda} = \frac{1-\lambda}{16\pi G_N}\int d^4x \sqrt{-g} \left(\nabla_{\mu}u^{\mu}\right)^2 \nonumber\\
 = \frac{1-\lambda}{16\pi G_N}\int d^4x \sqrt{-g} \left(2\varphi \nabla_{\mu}u^{\mu}-\varphi^2\right)\nonumber\\
 = \frac{1-\lambda}{16\pi G_N}\int d^4x \sqrt{-g} \left(-2u^{\mu}\partial_{\mu}\varphi -\varphi^2\right)\label{s_lambda}
\eea
where we introduce the non-dynamical auxiliary field $\varphi$, and used integration by parts to put $\nabla_{\mu}$ on $\varphi$ in the first term \footnote{A more systematic derivation of a covariant version of Ho${\rm \check{r}}$ava-Lifshitz theory can be found in \cite{Germani:2009yt}.}.

Now, let us turn to the quadratic cuscuton action \cite{Afshordi:2006ad}:
\beq
S_{qc} = \int d^4x \sqrt{-g} \left(\mu^2\sqrt{\partial^\nu\varphi\partial_\nu\varphi}-\frac{1}{2}m^2\varphi^2\right),
\eeq
where $\mu$ and $m$ are constants of the theory. This can be easily generalized to general potentials by replacing $m^2\varphi^2/2$ with an arbitrary $V(\varphi)$. Introducing the auxiliary fields $v^{\nu}$ and $\sigma$, we can re-write this action as:
\beq
S_{qc} = \int d^4x \sqrt{-g} \left[\mu^2v^\nu\partial_{\nu}\varphi-\frac{1}{2}m^2\varphi^2 +\sigma(v^\nu v_\nu-1)\right].\label{s_qc}
\eeq

We thus see that the low-energy Ho${\rm \check{r}}$ava-Lifshitz and quadratic cuscuton actions are the same, $S_{\lambda}=S_{qc}$  if:
\bea
\mu^2 = -m^2 = -\frac{1-\lambda}{8\pi G_N},\\
v^{\mu} = u^\mu =   \frac{\partial^{\mu}\psi}{\sqrt{\partial^{\nu}\psi\partial_{\nu}\psi}}.\label{vu}
\eea

However, there is a subtlety in this equivalence: the gradient (or irrotational) condition of Eq. (\ref{vu}) which ensures that the vector field threads a space-time foliation is enforced at the action level in the Ho${\rm \check{r}}$ava-Lifshitz theory (Eq. \ref{s_lambda}), but it is only realized at the level of equations of motion that follow from Eq. (\ref{s_qc}). As a result, extremizing $S_{qc}$ is a sufficient (but not necessary) condition to extremize $S_{\lambda}$. Nevertheless, since the vector field $v^\mu$ is non-dynamical, one expects the solutions to be the same with appropriate boundary conditions.

To see this explicitly, we can compare the equations of motion, while varying actions (\ref{s_lambda}) and (\ref{s_qc}) with respect to the vector fields. For $S_{qc}$, varying with respect $v^\mu$ yields:
\beq
v_{\mu} \propto \partial_\mu\varphi,
\eeq
which, given that $v^\mu$ has unit norm, fixes it completely.
On the other hand, varying $S_{\lambda}$ with respect to $\psi$ yields:
\beq
\partial_{\mu}\left[\sqrt{-g} \left(\partial^{\alpha}\psi\partial_{\alpha}\psi\right)^{-1/2}\left(g^{\mu\nu}-u^\mu u^\nu\right)\partial_\nu\varphi\right] =0,\label{eq_lambda}
\eeq
where $u_\mu \propto \partial_\mu\psi$. While $u_\mu \propto \partial_\mu\varphi$ solves this equation, it is clearly not the only possible solution. However, notice that if we use the Ho${\rm \check{r}}$ava-Lifshitz preferred reference frame (or uniform $\psi$ gauge), Eq. (\ref{eq_lambda}) reduces to a linear 2nd order elliptical equation for $\varphi$ with only spatial derivatives:
\beq
\partial_{i}\left(M^{ij}\partial_j\varphi\right) =0,\label{eq_lambda_i}
\eeq
 if we set $g_{0i}=0$, where
\beq
M^{ij}= \sqrt{-g} \left(\partial^{\alpha}\psi\partial_{\alpha}\psi\right)^{-1/2}g^{ij}.
\eeq
Similar to the Laplace equation, the solutions to $\varphi$ are thus fixed by boundary conditions at spatial infinity. In particular, a uniform $\varphi$ [or equivalently  $\varphi= f(\psi)$] at spatial infinity leads to completely uniform $\varphi$ on constant $\psi$ surfaces. Since, this constraint does not involve any time derivatives, it can be equally enforced at the classical and quantum level.

To summarize, as long as the constraint $u_{\mu} \propto \partial_\mu\varphi =  \partial_\mu \nabla_{\nu} u^\nu$ \footnote{Note that $\varphi = \nabla_{\nu} u^\nu$ follows from varying $S_{\lambda}$ with respect to $\varphi$.} is enforced at spatial infinity, the quadratic cuscuton and low energy Ho${\rm \check{r}}$ava-Lifshitz theories are equivalent. We should point out that this is naturally realized for solutions that asymptote to FRW cosmological geometries at spatial infinity. Now, varying $S_{qc}$ with respect to $\varphi$ , we recover the equation of motion for low energy Ho${\rm \check{r}}$ava-Lifshitz/cuscuton theories:
\beq
K= \frac{1}{\sqrt{-g}} \partial_\mu \left(\sqrt{-g}\partial^\mu\varphi\over\sqrt{\partial^\nu\varphi\partial_\nu\varphi}\right) = -\frac{m^2 \varphi}{\mu^2}\label{qc_eom}
\eeq

\section{CMC gauge and Cosmology}
We should note that the condition $u_{\mu} \propto  \partial_\mu \nabla_{\nu} u^\nu$ for normal vectors to constant $\psi$ surfaces implies that the mean extrinsic curvature $K = \nabla_{\nu} u^\nu$ is uniform on these surfaces. Therefore, as long as $\lambda \neq 1$, the preferred gauge in the low energy Ho${\rm \check{r}}$ava-Lifshitz (and cuscuton) theories are constant mean curvature (CMC) surfaces. Since $K=\nabla_{\nu} u^\nu$ is also the local expansion rate of the cuscuton fluid, this is also known as the uniform expansion gauge.

Using this equivalence, we can further use the cosmological constraints on the quadratic cuscuton model to constrain the parameter $\lambda$.  It was shown in \cite{Afshordi:2006ad,Afshordi:2007yx} that the quadratic cuscuton model causes an effective mis-match between the infrared (super-horizon) and ultraviolet (sub-horizon) Planck masses:
\bea
M^2_{p,{\rm IR}} = M^2_{p,{\rm UV}} - \frac{3\mu^4}{2m^2} = \left[1+\frac{3}{2}(1-\lambda)\right]M^2_{p,{\rm UV}},\nonumber\\
\eea
where $M_{p,{\rm UV}} = (8\pi G_{N})^{-1/2}$ is the reduced Planck mass on sub-horizon scales. This mis-match causes anomalous growth/decay of perturbations during the matter era, which can be tested by comparing large scale structure with the cosmic microwave background (CMB) anisotropies, or through the early Integrated Sachs-Wolfe effect in the CMB. \cite{Robbers:2007ca} finds the cosmological constraints on this phenomenon for the quadratic cuscuton model, which yields:
\beq
\lambda -1 = 0.003 \pm 0.014 ~~{\rm (95\%~CL)}.
\eeq
A weaker version of this constraint, based on big bang nucleosynthesis predictions for primordial abundances, was already found in \cite{Kehagias:2009is}.
\section{Geometric symmetries and lack of dynamics}

\begin{figure}
\includegraphics[width=1.3\linewidth]{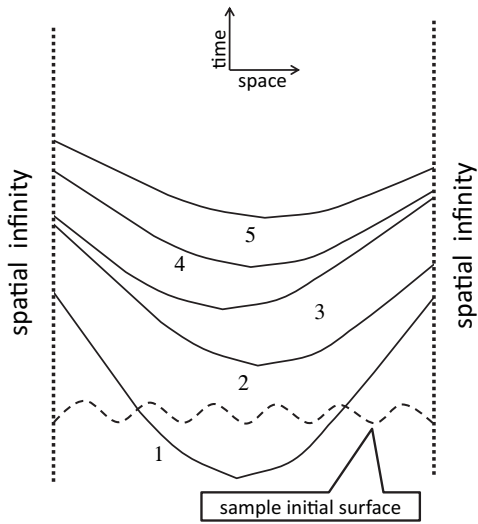}
 \caption{A cartoon of constant field hypersurfaces of cuscuton in a fixed space-time. As we argue in the text, different surfaces are decoupled in the action, and are thus fixed by boundary conditions at spatial infinity.}
\label{cartoon}
\end{figure}

The infinite speed of sound for the cuscuton theory implies that the linear perturbations of the field around any uniform background: $\pi(x^i,t)=\varphi(x^i,t)-\bar{\varphi}(t)$, coupled to external fields $J(x^i,t)$ (such as metric), satisfy a non-dynamical equation (e.g. \cite{Gao:2009ht}) of the form:
\beq
g^{ij}\nabla_{i}\nabla_{j}\pi + {\cal O}(\pi^2)= J(x^i,t),
\eeq
Therefore, $\pi(x^i,t)$ can be solved as a (non-local but simultaneous) function of $J(x^i,t)$ (and spatial boundary conditions) perturbatively, to any order, independent of initial conditions. In particular, no (ghost or tachyonic) dynamical instability can exist for the $\pi$ field, at least at the perturbative level.

The case of cuscuton in $1+1$ dimensional Minkowski space is particularly instructive, as Lorentzian CMC surfaces are hyperbolae. Therefore, any 1-d family of non-intersecting hyperbolae can describe the constant-$\varphi$ curves in a classical solution to the cuscuton field equation, where $\varphi$ is the extrinsic curvature of the hyperbola. A cartoon of this is shown in Fig. (\ref{cartoon}) which shows that fixing the two ends of a constant $\varphi$-surface (say at spatial infinities) fixes the entire surface.

Is there a well-defined initial value problem for the cuscuton field theory? As we argued above, any fluctuations around uniform $\varphi(t)$ backgrounds remain non-dynamical in perturbation theory, order by order. Nevertheless, for a general space-time foliation, the equation of motion (\ref{qc_eom}) has 2nd order time derivatives, and thus appears to be dynamical. This is only an illusion, as for any initial surface [e.g. the dashed line in Fig. (\ref{cartoon})], there is a non-local constraint that relates $\varphi$ and $\dot{\varphi}$ at two different points on the surface. In other words, fixing initial $\varphi$ and $\dot{\varphi}$ at one point in space fixes the constant-$\varphi$ hyperbola, which in turn sets these values where the hyperbola re-intersects the initial surface.

What we conclude from this discussion is that cuscuton initial conditions on a general initial surface should satisfy a (generally) non-local constraint, in order to lead to a consistent future evolution (at least for a finite time). Therefore, the theory is still globally non-dynamical, even though it may locally appear dynamical in a general gauge. In other words, a local canonical formulation of field theory is non-existent for this theory, which manifests itself as a degenerate (i.e. zero-volume) canonical phase space of linear fluctuations around constant field surfaces.
We should point out that neglecting this non-local constraint has led others to infer instabilities/ghosts, or inconsistencies in the Ho${\rm \check{r}}$ava-Lifshitz theory (e.g. \cite{Li:2009bg,Blas:2009yd,Kocharyan:2009te}). However, neglecting the gravitational backreaction, a classical solution to the scalar field equation is as simple as finding the CMC foliation of the spacetime, which is a generally well-defined procedure. We should note that after solving for the scalar constraint, the remaining metric degrees of freedom will obey a non-local dynamical equation of motion.

Another source of confusion has been linear perturbation theory around (flat foliation of) Minkwoski space (e.g. \cite{Horava:2009uw,Wang:2009yz,Charmousis:2009tc,Bogdanos:2009uj}). Flat foliation of Minkowski space is {\it not} a classical background solution to the cuscuton equation of motion, which invalidates any perturbation theory around it for $\lambda \neq 0$. This is similar to the case of degenerate perturbation theory in quantum mechanics, as the $\lambda-1$ term (as well as the higher derivative terms) breaks the foliation independence of the Einstein-Hilbert action.

We will next introduce a geometric symmetry principle that can protect this non-dynamical nature in the UV completion of cuscuton/Ho${\rm \check{r}}$ava-Lifshitz theories. Let us consider the infinitesimal field transformations:
\beq
\delta\varphi = a^\mu \partial_\mu f(\varphi),\label{dphi}
\eeq
where $f(\varphi)$ is an arbitrary function of $\varphi$ and $a^\mu$ is a vector field that satisfies:
\beq
\nabla^\mu a_\mu = \partial_\mu \varphi\partial_\nu \varphi \nabla^\mu a^\nu  = 0,\label{amu}
\eeq
i.e. the divergence of $a^\mu$, as well as the divergence of its projection onto constant $\varphi$ surfaces vanish.
The change in the quadratic cuscuton action, after some integration by parts, takes the form:

\bea
\delta S_{qc} = \int d^4x \sqrt{-g} \times \nonumber\\ \left[\frac{\mu^2 f'(\varphi) \nabla^\mu a^\nu\left(\partial_\mu \varphi\partial_\nu \varphi -  g_{\mu\nu} \partial^\alpha\varphi\partial_\alpha\varphi\right)}{\sqrt{\partial^\alpha\varphi\partial_\alpha\varphi}}  + h(\varphi) \nabla^\mu a_\mu \right],
\eea
which vanishes up to surface terms if we use (\ref{amu}), where
\beq
 h(\varphi) = m^2 \int f'(\varphi)\varphi d\varphi.
\eeq
Therefore, the cuscuton action is invariant under the transformations (\ref{dphi}) for any $f(\varphi)$, and a large class of $a^\mu$'s (e.g. any constant $a^\mu$ in a Minkowski space-time). These symmetries are unique to the cuscuton (and by extension to Ho${\rm \check{r}}$ava-Lifshitz) action(s), and can protect the low energy theory from higher derivative terms that could make $\varphi$ dynamical.

There is a deeper geometric meaning to the symmetries (\ref{dphi}), which are equivalent to coordinate transformations:
\beq
\delta x^{\mu} = a^{\mu} f'(\varphi).
\eeq
The fact that $f'(\varphi)$ is an arbitrary function of $\varphi$ is a consequence of the fact that both actions are a sum over the geometric functions of constant $\varphi$ surfaces:
\beq
S_{qc,HL} = S_{EH} + \int d\varphi~ \Omega_{qc,HL} [x^\mu(\sigma_i,\varphi)],
\eeq
where $x^\mu(\sigma^i,\varphi)$ characterize the constant $\varphi$ surfaces with a spatial parametrization $\sigma^i$. For example, $\Omega_{qc}$ is a linear combination of area (or 3-volume) $\Sigma(\varphi)$ and 4-volume ${\cal V}(\varphi)$ of the constant $\varphi$ surface \cite{Afshordi:2006ad}:
\bea
\Omega_{qc}[x^\mu(\sigma_i,\varphi)] = \mu^2 {\rm sgn}(\dot{\varphi}) \Sigma(\varphi) +m^2 \varphi {\cal V}(\varphi) \nonumber\\ = \frac{\lambda-1}{8\pi G_N} \left[ {\rm sgn}(\dot{\varphi}) \Sigma(\varphi) +\varphi {\cal V}(\varphi)\right], \label{Omega}
\eea
where $m^2\varphi$ can be replaced by $V'(\varphi)$ for a more general potential.  Most significantly, there is no coupling between different $\varphi$ surfaces (for fixed background metric), which, as we discussed above, implies that minimizing action fixes each surface for a given metric and given boundary conditions at spatial infinity, {\it independent of other surfaces or initial conditions of $\varphi$}.

The only exception to this statement is when different surfaces intersect (i.e. caustics form), which will signify a breakdown of the low energy theory. One may speculate that gravitational backreaction would prevent these singularities in a UV-complete theory of quantum gravity. At the classical level, it was shown in \cite{Afshordi:2007yx} that cosmological perturbations in $\varphi$ remain small (or identically the uniform expansion/CMC gauge remains well-defined), with possible exception of the vicinity of black hole horizons.

Notice that there is also a close analogy between Eq. (\ref{Omega}) and the energy of soap bubbles/films, with the area and volume terms characterizing the surface tension and pressure difference between the two sides of the film, respectively \cite{Afshordi:2006ad}. This naturally explains the emergence of CMC surfaces as the extremum of the action.

 Let us now consider the non-perturbative stability of the system. Even though $\varphi$ is non-dynamical, the space-time metric has dynamical degrees of freedom, which have a modified behavior due to coupling to cuscuton.
While the GR Hamiltonian is non-negative \cite{PhysRevD.10.2345}, the low energy Hamiltonian for the scalar field in an arbitrary coordinate system (but setting $g_{0i}=0$) has the form:
\beq
H_{\lambda} = \frac{\lambda-1}{8\pi G_N}\int d^3x \sqrt{-g} \left\{{|g^{ij}\partial_i\varphi\partial_j\varphi|\over \sqrt{\partial_\mu\varphi\partial^\mu\varphi}}-\frac{1}{2}\varphi^2\right\}.
\eeq
Although this is not a generic feature of cuscuton theories, we notice that the Ho${\rm \check{r}}$ava-Lifshitz Hamiltonian (with $\mu^2=-m^2$) is not positive definite. Therefore, coupling to positive energy sectors can potentially lead to non-perturbative instabilities. While $\lambda >1$ appears to yield a benign tachyonic negative energy, for $\lambda <1$ short wave length perturbations can have arbitrarily negative energies, which may signal a potential instability (also see \cite{Bogdanos:2009uj}), although the non-dynamical nature of $\varphi$ prevents a ghost-like degree of freedom. We should also point out that since this theory does not have an asymptotically flat space-time, the notion of energy/Hamiltonian may not be very helpful for analyzing the global stability of the system.

Finally, we should point out that the decoupling of different $\varphi$-surfaces can be broken by the so-called projectibility condition, that forces the proper time to advance uniformly on constant-$\varphi$ surfaces, or the Ho${\rm \check{r}}$ava-Lifshitz preferred frame. This constraint couples different surfaces, and thus can make the theory dynamical, leading to caustics and possibly ghost/tachynoic instabilities (e.g. \cite{Blas:2009yd,Kobakhidze:2009zr}, but also see \cite{Mukohyama:2009tp}).

\section{Conclusions}

We have shown that at low energies, Ho${\rm \check{r}}$ava-Lifshitz theory of gravity is identical to Einstein's gravity plus cuscuton, a non-dynamical/incompressible scalar field theory. The Lorentz breaking of the low energy theory is constrained to less than $1.4\%$, based on cosmological observations. The non-dynamical nature of the scalar is maintained at the non-perturbative level, which stems from decoupling of constant $\varphi$ surfaces in the cuscuton action, which breaks into sum of areas and volumes of these surfaces. Including higher order geometrical functionals of these surfaces (such as integrals of 3-curvature and its covariant derivatives), as proposed in the UV regime of the theory, does not change this decoupling, and thus the non-dynamical nature of the scalar sector. Nevertheless, coupling to a positive energy sector might lead to a non-perturbative UV instability for $\lambda < 1$.

Another important result of this geometrical picture is the natural emergence of CMC surfaces as the preferred frame of Ho${\rm \check{r}}$ava-Lifshitz gravity at low energies. It is interesting to notice a possible connection with the emergence of the CMC gauge in the timeless formulation of spatially conformal gravity in \cite{Anderson:2004bq}.

I am grateful to Bruce Bassett, Robert Brandenberger, Xiao Liu, Constantinos Skordis, Herman Verlinde, and Tom Zlosnik for useful discussions. I am particularly indebted to Andrew Tolley for many discussions through the course of this project.
NA is supported by Perimeter
Institute (PI) for Theoretical Physics.  Research at PI is
supported by the Government of Canada through Industry Canada and by the
Province of Ontario through the Ministry of Research \& Innovation.

\vspace*{-0.5cm}

\bibliographystyle{utcaps_na2}
\bibliography{horava_cuscuton}

\end{document}